\documentclass[reprint,aps, prl,superscriptaddress]{revtex4-2}

\usepackage{amsmath, amssymb, times, mathrsfs, hyperref, array, bbm}
\usepackage{graphicx}
\usepackage[usenames,dvipsnames]{xcolor}
\usepackage{braket}

\hypersetup{colorlinks=false}

\bibpunct{[}{]}{,}{n}{}{}
\usepackage{ulem}

\begin{document}
\title{Parity protected superconducting diode effect in topological Josephson junctions}

\author{Henry F. Legg}
\affiliation{Department of Physics, University of Basel, Klingelbergstrasse 82, CH-4056 Basel, Switzerland}

\author{Katharina Laubscher}
\affiliation{Department of Physics, University of Basel, Klingelbergstrasse 82, CH-4056 Basel, Switzerland}
\affiliation{Condensed Matter Theory Center and Joint Quantum Institute, Department of Physics, University of Maryland, College Park, Maryland 20742, USA}

\author{Daniel Loss}
\affiliation{Department of Physics, University of Basel, Klingelbergstrasse 82, CH-4056 Basel, Switzerland}

\author{Jelena Klinovaja}
\affiliation{Department of Physics, University of Basel, Klingelbergstrasse 82, CH-4056 Basel, Switzerland}

\begin{abstract}
In bulk superconductors or Josephson junctions formed in materials with spin-orbit interaction, the critical current can depend on the direction of current flow and applied magnetic field, an effect known as the superconducting (SC) diode effect. Here, we consider the SC diode effect in Josephson junctions in nanowire devices. We find that the $4\pi$-periodic contribution of Majorana bound states (MBSs) to the current phase relation (CPR) of single junctions results in a significant enhancement of the SC diode effect when the device enters the topological phase. Crucially, this enhancement of the SC diode effect is independent of the parity of the junction and therefore protected from parity altering events, such as quasiparticle poisoning, which have hampered efforts to directly observe the $4\pi$-periodic CPR of MBSs. We show that this effect can be generalized to SQUIDs and that, in such devices, the parity-protected SC diode effect can provide a highly controllable probe of the topology in a Josephson junction. 
\end{abstract}

\maketitle

{\it Introduction.} The topological protection of MBSs makes them promising candidates for fault tolerant topological quantum computing \cite{Ivanov2001,Kitaev2003,Nayak2008,Alicea2012}. MBSs are predicted to appear in the cores of vortices or at boundaries of topological superconductors \cite{Volovik1999,Read2000,Fu2008}. One of the most mature device architectures purported to host MBSs consists of a nanowire brought into proximity with a superconductor, such as Al or Nb \cite{Laubscher2021}. In such devices, the nanowire, for instance, can be a semiconducting nanowire with strong Rashba spin-orbit interaction (SOI) \cite{Oreg2010,Lutchyn2010} or a topological insulator nanowire \cite{Cook2011,Legg2021}. Although signatures consistent with MBSs, such as zero-bias peaks in local conductance measurements \cite{Sasaki2011, Mourik2012, Deng2012,Das2012, Churchill2013,Prada2020}, have been observed, these signals can also be explained by trivial Andreev bound states (ABSs) and there has been no conclusive experimental observation of MBSs so far \cite{Andreev1966,Kells2012,Lee2012, Cayao2015, Reeg2018, Penaranda2018, Vuik2019, Woods2019, Liu2019,Chen2019, Awoga2019, Valentini2020, Prada2020, Hess2021}. 

The lack of a conclusive experimental observation has led to the suggestion of several auxiliary features in nanowire devices that could indicate the onset of a topological phase. Examples include looking for correlated zero-bias peaks, oscillations around zero energy due to a finite overlap of the MBSs in short nanowires \cite{Prada2012,DS2012,Rainis2013, Dmytruk2018, Fleckenstein2018}, the flip of the spin polarization in the lowest band \cite{Szumniak2017,Chevallier2018}, a quantized conductance peak with height $2e^2/h$  \cite{Law2009, Akhmerov2009,Flensberg2010,Wimmer2011,Chevallier2016}, and using non-local conductance to observe the bulk gap closing and reopening associated with the topological phase transition \cite{Entin2008,Lobos2014,Gramich2017,Rosdahl2018,Zhang2019,Danon2020,Melo2021, Hess2021,Pan2021,Pikulin2021,Microsoft2022,Hess2022}. 

\begin{figure}[t]
\includegraphics[width=0.75\columnwidth]{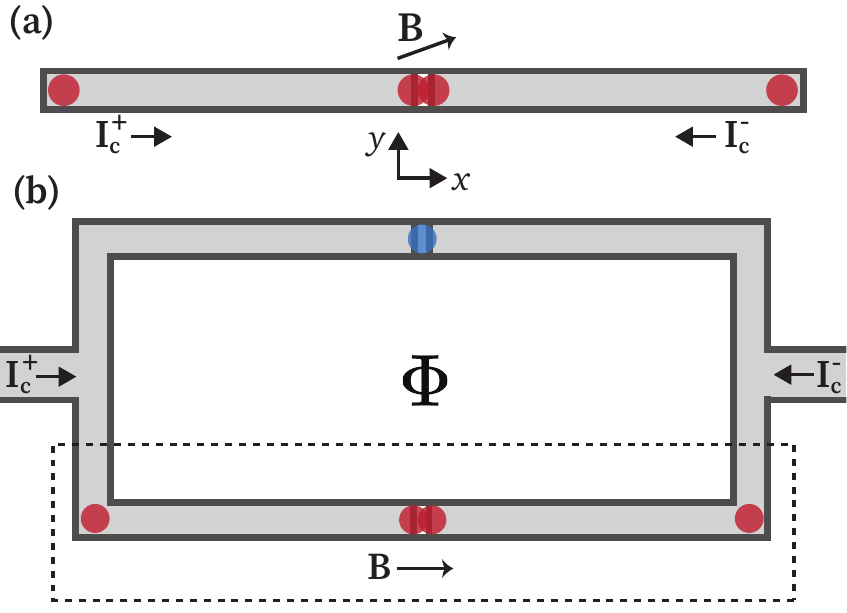}
\caption{\linespread{1.02}\selectfont{}
{\it Setups to measure the parity-protected SC diode effect in topological Josephson junctions.} {\bf (a)} In a single junction the SC diode effect, meaning that the critical current is directional dependent  $I_c^+\neq I_c^-$, can occur when the applied magnetic field has a component parallel to the SOI vector (here $y$-direction). If, in addition, a component of the magnetic field perpendicular to the SOI vector drives the system into the topological SC phase, then MBSs form both in the junction and at the ends of the nanowire (red dots). The $4\pi$-periodic CPR of the MBSs in the junction results in a significant enhancement of the SC diode effect that is independent of the parity of the junction. {\bf (b)} The SC diode effect can also be probed using a SQUID that consists of a topologically trivial low transparency junction that hosts only ABSs (blue dot) and a junction that can be driven into the topological SC phase hosting MBSs (red dots). At finite flux $\Phi$, a large parity-protected SC diode effect occurs when the junction is in the topological phase.}\label{fig1}
\end{figure}

In Josephson junctions (JJs), see Fig.~\ref{fig1}, formed by a weak link in the superconductor, another signature that initially appeared promising was the $4\pi$-periodic current phase relation (CPR) that is expected when two MBSs hybridize with each other in the junction \cite{Law2011,Jose2012,Dominguez2012,Houzet2013,Peng2016,Dominguez2017,Pico2017}. However, this signal turned out to be highly sensitive to parity altering events such as quasiparticle poisoning which renders the CPR in experimental systems $2\pi$-periodic \cite{Law2011,Jose2012,Dominguez2012,Houzet2013,Peng2016,Dominguez2017,Pico2017,Spanton2017,Cayao2017,Cayao2018,Kayyalha2020,Svetogorov2022}. Whilst missing odd Shapiro steps in JJs under AC bias might hint at the presence of topological superconductivity and associated MBSs \cite{Wiedenmann2016,Bocquillon2017,Laroche2019,Rosenbach2021,Fischer2022}, these observations can also occur due to trivial mechanisms such as Landau-Zener transitions \cite{Chiu2019,Dartiailh2021,Frolov2022}. 

Another phenomenon that can occur in materials with strong SOI is the so-called SC diode effect \cite{Fulton1972,Wakatsuki2017,Ando2020,Kononov2020,Baumgartner2021,Daido2022,He2022,Hou2022,Noah2022,Ilic2022,Wu2022,Legg2022,Baumgartner2022,Fominov2022,Souto2022,Mazur2022,Zhang2022,Davydova2022,Pal2022,Bauriedl2022,Trahms2022,YZhang2022}. In particular, magnetochiral anisotropy of the energy spectrum can result in a critical current of a material or junction that is dependent on the direction of current flow and applied magnetic field \cite{Baumgartner2021,Legg2022,Legg2022MCA}. This results in a range of current for which supercurrent can flow only in one direction through the material or junction. In the bulk of proximitized nanowires this effect can be used, e.g. to determine the direction of the SOI vector and provide a measure of the strength of the SOI in the presence of a superconductor \cite{Legg2022,Zhang2022,Legg2022MCA}.

In this paper we consider the SC diode effect in JJs in nanowire devices \cite{Spanton2017,Mazur2022,Zhang2022}. First, we show that the $4\pi$-periodic contribution of MBSs to the CPR results in a significant enhancement of the SC diode effect when a single junction, as shown in Fig.~\ref{fig1}(a), enters a topological SC phase. Importantly, whilst the full CPR does depend on the parity of the junction, the size and sign of the SC diode effect is robust against parity altering events, such as quasiparticle poisoning, and therefore is a parity protected signal of the MBSs in the junction.  We then generalize this effect to  SC interference devices (SQUIDs) where the parity-protected SC diode effect  provides a highly controllable probe of the topology of a junction. Finally, we discuss the experimental implementation of these two related signatures of MBSs.

{\it CPRs of single junctions.} 
The energy of a JJ with phase difference $\varphi$ across the junction contains contributions from occupied bound and continuum states \cite{Svetogorov2022}
\begin{equation}
E(\varphi)=\sum_n \varepsilon_A^{(n)}(\varphi) +  \xi \varepsilon_M(\varphi) + \varepsilon_c(\varphi),
\end{equation}
where $\varepsilon_A^{(n)}(\varphi)<0$ and  $\varepsilon_M(\varphi)$ are the energies of the ABSs and (hybridized) MBSs in the junction, respectively, and $\xi=\pm 1$ governs the parity of the MBSs. Here, $\varepsilon_c(\varphi)<0$ captures the phase dependence of the continuum states, however, since these have the same period in $\varphi$ as ABSs, we will first focus on the contributions only from bound states and numerically analyze the full system later.

The energies of ABSs are $2\pi$-periodic in $\varphi$, such that in the general case we can write the energy in terms of a Fourier series \cite{Spanton2017,Baumgartner2022,Fominov2022}
\begin{equation}
\varepsilon_A^{(n)}(\varphi)=A_0 +\sum_m \frac{A_m^{(n)}}{m} \cos(m \varphi+\phi_m^{(n)}),
\end{equation}
here $m=1,2,\dots$ label the corresponding CPR harmonics which are governed by the transparency of the junction. Similarly for the $4\pi$-periodic MBSs the energy can be written as
\begin{equation}
\varepsilon_M(\varphi)=M_{0}+\sum_{m} \frac{2 M_{m}}{m} \cos(m \varphi/2+\phi_m^M),\label{MBScpr}
\end{equation}
here $m=1,2,\dots$ labels the half-integer harmonics. In both cases we also include phases, e.g. $\phi^{(n)}_m$, these phases can be non-trivial when time-reversal symmetry and inversion symmetry are broken \cite{Szombati2016,Schrade2017}, e.g. as in Fig.~\ref{fig1}(a). The full CPR for the junction with ground state energy $E(\varphi)$ (at zero temperature) is then given by
\begin{align}
I(\varphi)&=-\frac{e}{h} \frac{\partial E (\varphi)}{\partial \varphi}=\sum_n I^A_n(\varphi) + I^M_{\xi}(\varphi),\label{fullCPR}
\end{align}
where $e<0$ is the electron charge, $I^A_n(\varphi)$ and $I_\xi^M(\varphi)$ are the individual CPR contributions of ABSs and the MBSs, respectively.

\begin{figure}[t]
\includegraphics[width=\columnwidth]{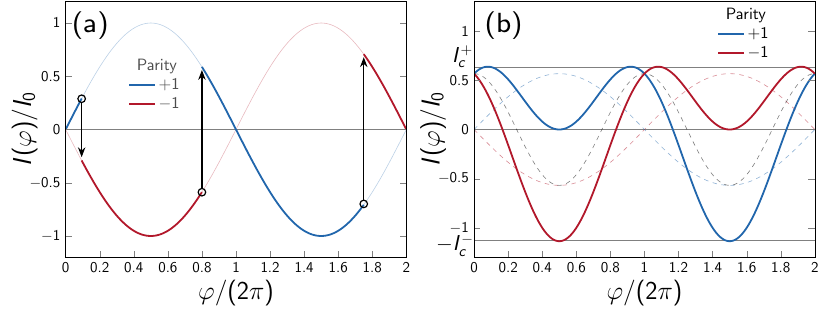}
\caption{\linespread{1.05}\selectfont{}
{\it Parity-protected SC diode effect in topological JJs.} {\bf (a)} Schematic showing the impact of parity altering events in CPR from MBSs. Since measurement of the full CPR cannot be done faster than the rate of parity altering events, such as quasiparticle poisoning, measurement of the theoretical $4\pi$-periodic CPR of the MBSs in a topological JJ is rendered effectively $2\pi$-periodic. {\bf (b)} The interference between the $2\pi$-periodic CPRs of ABSs and the $4\pi$-periodic CPR of MBSs, here both of amplitude $I_0$, can lead to a significant SC diode effect in both a single junction and SQUID (see Fig.~\ref{fig1}). The full CPR remains parity dependent (indicated by red and blue lines, individual bound state CPR contributions are indicated by dashed lines). In contrast, however, the size and sign of the SC diode effect, governed by $\Delta I=I^+_c-I_c^-$, is protected from parity altering events and can therefore be used as a measure of junction topology.}\label{fig2}
\end{figure}

 \begin{figure*}[t]
\includegraphics[width=2\columnwidth]{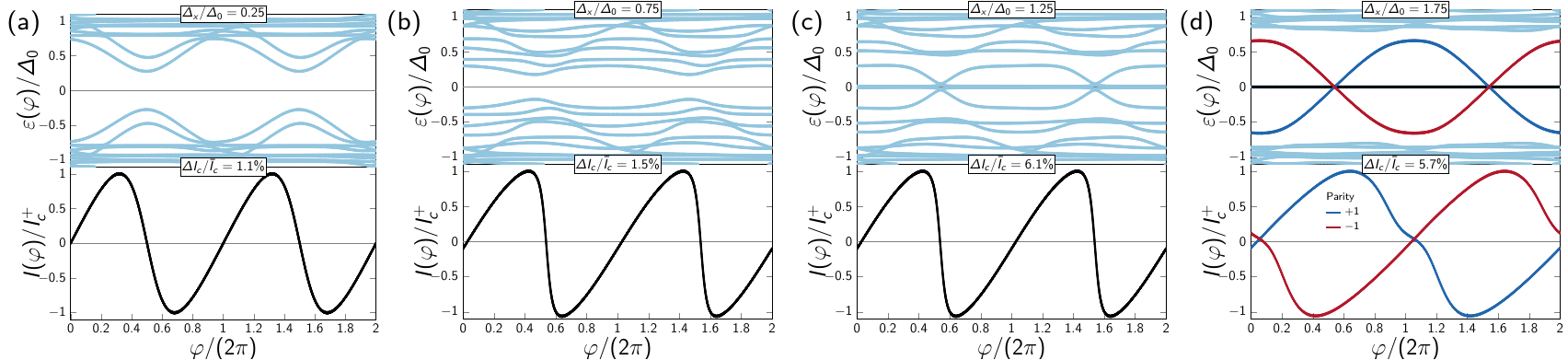}
\caption{\linespread{1.0}\selectfont{}
{\it Parity-protected SC diode effect in a single nanowire JJ with both parallel ($\Delta_y$) and perpendicular ($\Delta_x$) Zeeman fields to the SOI vector.}  {\bf Top panels:} energy phase relations, $\varepsilon(\varphi)$, of ABSs and continuum states (light blue), interior MBSs (red/blue), and exterior MBSs (black: at zero energy and independent of $\varphi$), over a $4\pi$ period as obtained from the tight-binding model in Eq.~\eqref{eq:tb}, for various perpendicular fields $\Delta_x$ and fixed parallel field $\Delta_y=0.025 \Delta_0$.  {\bf Bottom panels:} Corresponding CPR, $I(\varphi)$, exhibiting an SC diode effect. The SC diode effect results from a difference in magnitude between the maxima and minima of the CPR, which determine the directional dependent critical currents, $I_c^\pm$. {\bf (a-b)} When the junction is in the trivial phase, $\Delta_x<\Delta_0$ for our parameters, only a weak SC diode effect is present, even for junctions with visible deviations from a purely sinusoidal CPR, see also Fig.~\ref{fig4}(a-b).  Note that, within our Rashba nanowire model, the SC diode effect is only present when $\Delta_x\neq0$, $\Delta_y\neq0$, and $\alpha\neq0$ are all satisfied, since these are necessary to manifest both broken inversion and time-reversal symmetry, see similar discussions in Ref.~\cite{Legg2022,Dantas2022}.  {\bf (c-d)} In the topological phase, $\Delta_x>\Delta_0$, MBSs begin to form and a much larger SC diode effect is present. In particular, when the exterior MBSs are well localized at the ends of the SC sections away from the junction, indicated by zero-energy states entirely independent of $\varphi$ (black), as in (d) but not in (c), the SC diode effect is independent of the parity of the MBSs. The diode efficiency is given by $\Delta I/\bar{I}_c$, where $\Delta I=I^+_c-I_c^-$ and $\bar{I}_c=(I^+_c+I_c^-)/2$. Parameters: $a=7.5$ nm, $N_s=200$, $N_b=3$, $N_j=5$, $t=6.8$ meV, $\alpha=3.33$ meV, $V=1$~meV, $\Delta_0=0.25$ meV.}\label{fig3}
\end{figure*}

{\it Parity-protected SC diode effect in a single junction.}
We now investigate the diode effect for a single junction in the setup shown in Fig.~\ref{fig1}(a). Throughout we will consider low transparency junctions, where higher harmonics of bound state energies can be neglected. In this case, the full CPR of a topologically trivial junction hosting multiple ABSs  is given by 
\begin{equation}
I_{\rm triv}(\varphi)\approx \sum_n A_1^{(n)} \sin(\varphi+\phi_1^{(n)})=\tilde{A} \sin (\varphi + \phi'),
\label{1JJ}
\end{equation}
where the coefficients satisfy $\tilde{A}\cos\phi'=\sum_n A^{(n)}_1 \cos \phi_1^{(n)}$ and  $\tilde{A}\sin\phi'=\sum_n A^{(n)}_1 \sin \phi_1^{(n)}$. We also define the (directional dependent) critical currents
\begin{equation}
I_c^+=\max \{I(\varphi)\}\;\;\;\;\&\;\;\;\;I_c^- =-\min\{I(\varphi)\}.
\end{equation}
For this low transparency trivial case, the purely sinusoidal CPR,  $I_{\rm triv}(\varphi)=\tilde{A} \sin (\varphi + \phi')$, still satisfies $I_{\rm triv}(\varphi)=-I_{\rm triv}(\varphi+\pi)$ ensuring that no diode effect is present, even if there is an assortment of finite phases $\phi^{(n)}_1$ for the individual ABS contributions to the CPR. In a more realistic trivial junction (see below) higher harmonics, e.g. $A_{2}^{(n)}$, will result in a finite diode effect \cite{Fulton1972,Baumgartner2022,Souto2022}. However, the SC diode effect resulting from higher harmonics will typically be small and, furthermore, can be easily suppressed in experiments by choosing junctions that have a low transparency.

We now consider the case where the junction is topological and therefore hosts a pair of MBSs. As discussed above, the CPR of the MBSs is $4\pi$-periodic and not robust against parity altering events which change the sign of the CPR contribution, see Fig.~\ref{fig2}(a). 
Again assuming low transparency, the full junction CPR is well approximated by
\begin{equation}
I_{\rm top}(\varphi)\approx \tilde{A} \sin (\varphi + \phi')+\xi M_1 \sin(\varphi/2+\phi^M_1).
\end{equation}
Unlike in the trivial phase, in the topological phase the contributions to the CPR of the ABS and the MBSs interfere and a finite SC diode effect can exist, $I_c^+\neq I_c^-$, even for low transparency junctions as long as $\phi' \neq 2 \phi^M_1$, see Fig.~\ref{fig2}(b). Furthermore, although in the topological state the full CPR is $4\pi$-periodic and not protected against parity changes, the directional dependent critical currents $I_c^+$ and $I_c^-$ are {\it independent} of the parity due to the relation $I^M_\xi (\varphi)=I^M_{-\xi} (\varphi+2\pi)$. This means that in a low transparency junction, where higher CPR harmonics are negligible, the SC diode effect is absent in the trivial state and finite in the topological state. Importantly, the size and sign of SC diode effect, governed by $I^+_c-I_c^-$, in the topological state is protected against parity altering events such as quasiparticle poisoning. Therefore, the SC diode effect can provide a parity independent probe of whether the junction is in the topological or trivial phase.

 \begin{figure*}[t]
\includegraphics[width=2.\columnwidth]{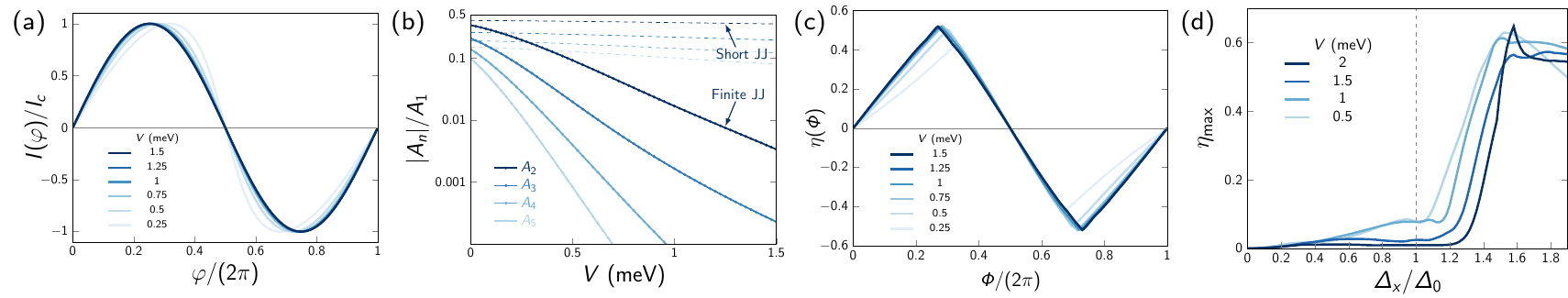}
\caption{\linespread{1.05}\selectfont{}
{\it CPRs of finite JJs and parity-protected SC diode effect in a SQUID.} {\bf (a)}  Dependence of CPR of a single JJ of finite length on barrier height, $V$, for $\Delta_x=0$. Note that the CPRs are normalised by $I_c$, which is also modified by $V$. {\bf (b)} Relative harmonics of the CPR in a single JJ as a function of barrier height $V$ for a finite JJ width ($N_j=15$) and short JJ ($N_j=1$) for $\Delta_x=0$. In a finite JJ the relative harmonics fall off considerably faster than in a short JJ, reducing the size of the SC diode effect in the trivial phase in both single JJs and SQUID for finite barrier heights. {\bf (c)} The SC diode figure of merit, $\eta(\Phi)$, as a function of the flux, $\Phi$, through the SQUID loop when one junction is in the topological phase ($\Delta_x=5\Delta_0/3$). For most transparencies the maximal $|\eta(\Phi)|$ occurs close to odd multiples of $\Phi/2$. For all $\Phi$, $\eta(\phi)$ is independent of the parity of the topological junction. {\bf (d)} Maximal absolute SC diode effect, $\eta_{\rm max}=\max\{|\eta(\Phi)|\}$, as a function of Zeeman field, $\Delta_x$, as one junction is tuned into the topological phase via application of a Zeeman field. In the trivial phase, $\Delta_x < \Delta_0$ for our parameters, the maximal $\eta(\Phi)$ is small for all barrier heights, $V$. However, in the topological phase, $\Delta_x >\Delta_0$ for our parameters (indicated by dashed gray line), the maximal $\eta(\Phi)$ increases substantially regardless of barrier height and is independent of the parity, therefore providing a measure of the onset of the topological phase. Parameters: $a=2.5$ nm, $N_s=480$, $N_b=6$, $N_j=15$, $t=61$ meV, $\alpha=5$ meV, $\Delta_y=0$, $\Delta_0=0.15$ meV.  For the SQUID (c) and (d) the trivial JJ is modeled using the same parameters but with $\Delta_x=0$. \label{fig4}}
\end{figure*}

{\it Numerical model.} So far our discussion has been analytic. We now show that the features discussed above are present in a more realistic numerical model of a JJ. Although the parity-protected SC diode effect is quite general and we would expect similar results from other nanowire systems, e.g. topological insulator nanowires \cite{Legg2021}, for simplicity we choose to model a JJ formed between two finite SC sections of a semiconductor nanowire with Rashba SOI. The tight-binding Hamiltonian describing this system is given by
\begin{align}
&H=\sum_{n,\nu,\nu'}\left\{c _ { n , \nu } ^ { \dagger }\left(-t \;\delta_{\nu \nu^{\prime}}+i \alpha\;\sigma_{\nu \nu^{\prime}}^y\right) c_{n+1, \nu^{\prime}} \right.+\frac{\Delta_n}{2} c_{n, \downarrow}^{\dagger} c_{n, \uparrow}^{\dagger} \nonumber \\
&+\frac{1}{2}c_{n, \nu}^{\dagger} \left[ \left( 2t-\mu_n\right) \delta_{\nu \nu^{\prime}}+\Delta_x \sigma_{\nu \nu^{\prime}}^x +\Delta_y \sigma_{\nu \nu^{\prime}}^y\right] c_{n, \nu^{\prime}} \left.+\text {H.c.}\right\},\label{eq:tb}
\end{align}
where $c_{n,\nu}^{\dagger}\, (c_{n,\nu})$ creates (annihilates) an electron with spin $\nu\in\{\uparrow,\downarrow\}$ at site $n$ in a chain of $N$ sites with lattice spacing $a$. Here, $n$ runs from 1 to $N=2N_s+N_j+N_b+1$, and $\nu,\nu'$ run over $\uparrow,\downarrow$, with $N_s$, $N_j$, $N_b$ the number of sites in each SC section, the JJ, and each barrier, respectively. The Pauli matrices $\sigma^l$, with $l \in \lbrace x,y,z \rbrace$, act in spin space with $y$ the direction along the SOI vector and $x$ the direction along the nanowire. For simplicity the hopping matrix element $t$, the Rashba SOI strength $ \alpha$, and  the Zeeman energies $\Delta_{y (x)}$ corresponding to Zeeman fields parallel (perpendicular) to the SOI vector are assumed to be constant throughout the system. On the other hand, the proximity induced SC gap $\Delta_n$ and the chemical potential $\mu_n$ depend on position, indexed by $n$. The pairing potential is finite only in the SC sections with a phase difference $\varphi$ across the JJ, such that the profile is given by
\begin{align}
\Delta_n =\Delta_0 \;{\rm if}\; n\leq N_s,\;\Delta_0 e^{i\varphi} \;{\rm if}\; N-n<N_s,
\end{align}
and is otherwise zero. Similarly within our model the chemical potential $\mu_n$ defines barriers of height $V$ on either side of the junction, such that \begin{equation}
\mu_n=V\;\;{\rm if}\;\;N_b^-< n\leq N_b^+\;{\rm or}\;\;N_b^- <N-n\leq N_b^+
\end{equation}
and is otherwise zero. Here we defined the barrier regions $N_b^{\pm}=N_s\pm N_b/2$. This profile means that the bulk chemical potential in the SC sections is located at the crossing point of the Rashba spin-split bands but the exact location of the chemical potential is not important, as long as the topological phase with well localized MBSs can be achieved. 

In Fig.~\ref{fig3} we show that the above analytic discussion of a single junction is an accurate representation of the features found using our more realistic numerical model, which includes all ABSs and continuum states, as well as MBSs if present. In particular, although higher harmonics result in a small SC diode effect in the trivial junction when $\Delta_x\neq0$, $\Delta_y\neq0$, and $\alpha\neq0$, there is a sizeable increase when the system enters the topological phase. Importantly, when the MBSs are well localized at the ends of the nanowire, such that the CPR is $4\pi$-periodic, we find that the size of the SC diode effect is independent of the parity, as shown in Fig.~\ref{fig3}(d). Note that the overlap of MBSs in the junction with exterior MBSs -- which occur at the outer ends of the SC section, see Fig.~\ref{fig1}(a) -- always results in a small finite energy of the lowest state $\varepsilon_0(\varphi)$. We neglect this finite energy when $\max\{\varepsilon_0(\varphi)\}=0.002\Delta_0$, assuming that below this threshold MBSs are sufficiently well localized such that they can be considered $4\pi$-periodic, however, the exact value of the threshold does not affect our results.

As discussed above, the SC diode effect in the trivial phase can be suppressed by choosing low transparency junctions. Within our numerical model the barrier height, $V$, controls the relative sizes of the harmonics $A_n$, see Fig.~\ref{fig4}(a-b). Furthermore, we want to emphasise that in a junction of finite length, $N_j\gg1$, the dependence of the relative magnitude of harmonics, $A_n/A_1$, on barrier height is much stronger than in short JJs \cite{Beenakker1991}, $N_j=1$, see Fig.~\ref{fig4}~(b). In fact, in all setups and for a large range of transparencies, we find the SC diode effect in the trivial phase is small and only becomes sizeable in the topological phase.

{\it Josephson diode interferometer.} We now investigate the parity-protected SC diode effect in SC interference devices (SQUIDs) as in Fig.~\ref{fig1}(b). We will see that this generalizes the case of a single junction and therefore provides a highly controlled probe of the topology of a junction. The setup consists of two junctions with CPRs $I_1(\varphi)$ and $I_2(\varphi)$ as well as of a flux $\Phi$, in units of the flux quantum $\Phi_0=h/2e$, that threads the SQUID loop. The CPR of the full system is given by the sum of the CPRs through the individual JJs~\cite{Fulton1972,Souto2022},
\begin{equation}
I(\varphi)=I_1(\varphi)+I_2(\varphi+\Phi).
\end{equation}
In particular, we see that this is similar to Eq.~\eqref{fullCPR}, but in the case of a SQUID the flux $\Phi$ naturally results in a phase difference between CPR contributions. As such, in this  setup no Zeeman field parallel to the SOI vector is required to obtain a diode effect and so we set the phases $\phi_m^{(n)}=\phi_m^M=0$. We define the SC diode figure of merit 
$\eta(\Phi)=(I_c^+-I_c^-)/I_{0}$,
where $I_{0}=I_{c}^\pm$ is the critical current for zero flux. When both junctions are purely sinusoidal and in the trivial phase, such that $I_1(\varphi)\propto I_2(\varphi)=A_1 \sin \varphi$, the full CPR satisfies the relation $I(\varphi+\pi)=-I(\varphi)$, ensuring that $\eta(\Phi)=0$ and no SC diode effect is present. As previously, higher harmonics can result in an SC diode effect in the trivial phase \cite{Fulton1972,Souto2022}, but this is small for low transparency junctions, see Fig.~\ref{fig4}(d).

We now consider when one junction of the SQUID is topological and the other is trivial. Similarly to a single junction, when the flux $\Phi\neq \pi n$ with $n\in \mathbb{Z}$, we find that $I_c^+\neq I_c^-$ and a large SC diode effect, $\eta(\Phi)\sim 60\%$, is possible, see Fig.~\ref{fig4}(c). The largest effect occurs for $\Phi\approx (2n+1)\pi/2$ with $n\in \mathbb{Z}$. Furthermore, as in the single junction, $\eta(\Phi)$ is independent of the parity of the topological junction and therefore the presence of an SC diode effect through a SQUID can be used to detect the onset of the topological SC phase, see Fig.~\ref{fig4}(d).

{\it Experimental considerations.} The parity-protected SC diode effect can be achieved with currently available experimental systems \cite{Spanton2017,Mazur2022,Zhang2022}, requiring only low transparency junctions that can be tuned into the topological phase, e.g., via electrostatic gating and a magnetic field. Furthermore, since the magnetic field configurations in the SQUID and single junction case are different, these two setups provide separate probes that can be performed in the same junction. Measurements of the parity-protected SC diode effect could be further augmented by other topological signatures, such as in local conductance measurements or missing Shapiro steps \cite{Prada2020,Jose2012,Houzet2013}, in order to provide further evidence that the system has entered the topological phase.

{\it Discussion.} We considered the SC diode effect in topological JJs. First, we showed that in single junctions the combination of a magnetic field parallel and perpendicular to the SOI vector results in a  SC diode effect that is substantially enhanced in the topological phase. Crucially, when MBSs are spatially well localized, then the size and sign of the SC diode effect is independent of the parity and therefore the SC diode effect is robust against e.g. quasiparticle poisoning. Furthermore, we showed that the parity-protected SC diode effect in topological junctions can be generalized to a SQUID geometry, where no Zeeman field parallel to the SOI vector is required. The discovery of the parity-protected SC diode effect opens up new ways to probe topological superconductivity.

\begin{acknowledgments}
{\it Acknowledgments.} This work was supported by the Georg H. Endress Foundation, the Swiss National Science Foundation, and NCCR QSIT (Grant number 51NF40-185902). This project received funding from the European Union’s Horizon 2020 research and innovation program (ERC Starting Grant, Grant No 757725). K. L. acknowledges support by the Laboratory for Physical Sciences through the Condensed Matter Theory Center.
\end{acknowledgments}
\bibliography{SC-diode.bib}

\end{document}